\documentclass[runningheads,a4paper]{llncs}

\usepackage{amsfonts,amssymb,amsmath}
\usepackage{natbib}

\usepackage{wrapfig}
\usepackage{llncsdoc}
\usepackage{epsfig,psfrag}
\usepackage{latexsym}
\usepackage{longtable}
\usepackage{tikz,pgf}
\usepackage{subfig}
\usepackage{wrapfig}

\usepackage{graphicx}
\usepackage{epstopdf}
\usepackage{color}

\newcommand{\bqn}{\begin{eqnarray}}
\newcommand{\eqn}{\end{eqnarray}}
\newcommand{\bq}{\begin{eqnarray*}}
\newcommand{\eq}{\end{eqnarray*}}
\usepackage[ruled,vlined]{algorithm2e}

\usepackage{url}
\usepackage{hyperref}
\hypersetup{colorlinks,%
citecolor=black,%
filecolor=blue,%
linkcolor=red,%
urlcolor=blue,%
pdftex}

\newcommand{\blue}[1]{{\color{blue} #1}}

\begin{document}

\title{Statistical Analysis on Brain Surfaces}
 \titlerunning{Statistical Analysis on Brain Surfaces}
 
\author{Moo K. Chung, Jamie L. Hanson, Seth D. Pollak
 }
\institute{
University of Wisconsin-Madison, USA\\
\vspace{0.3cm}
\blue{\tt mkchung@wisc.edu}
}
\authorrunning{Chung}

\maketitle


\begin{abstract}
\index{cortical surface analysis}

In this paper, we review widely used statistical analysis frameworks for data defined along cortical and subcortical surfaces that have been developed in last two decades. The cerebral cortex has the topology of a 2D highly convoluted sheet. For data obtained along curved non-Euclidean surfaces, traditional statistical analysis and smoothing techniques based on the Euclidean metric structure are inefficient. To increase the signal-to-noise ratio (SNR) and to boost the sensitivity of the analysis, it is necessary to smooth out noisy surface data.  However, this requires smoothing data on curved cortical manifolds and assigning smoothing weights based on the geodesic distance along the surface. Thus, many cortical surface data analysis frameworks are differential geometric in nature \citep{chung.2012.CNA}. The smoothed surface data is then treated as smooth random fields and statistical inferences can be performed within Keith Worsley's random field theory \citep{worsley.1996,worsley.1999}. The methods described in this paper are illustrated with the hippocampus surface data set published in \citep{chung.2011.hbm}. Using this case study, we will determine if there is an effect of family income on the growth of hippocampus in children in detail. There are a total of 124 children and 82 of them have repeat magnetic resonance images (MRI) two years later. 
\end{abstract}

\section{Introduction}

The cerebral cortex has the topology of a 2D convoluted
sheet.  Most of the features that distinguish these cortical
regions can only be measured relative to the local orientation of
the cortical surface \citep{dale.1999}. As the brain develops
over time, cortical surface area expands and  its curvature changes \citep{chung.2012.CNA}. 
It is equally likely that such
age-related changes with respect to the cortical surface are not
uniform \citep{chung.2003.NI,thompson.2000}. By measuring how geometric features such as the
cortical thickness, curvature and local surface area change over
time, statistically significant brain tissue growth or loss in
the cortex can be detected locally at the vertex level.

The first obstacle in performing surface-based data analysis is a need for extracting 
cortical surfaces from MRI volumes. This requires correcting MRI field inhomogeneity artifacts. The most widely used technique is the nonparametric nonuniform intensity normalization method (N3) developed at the Montreal neurological institute (MNI), which eliminates the dependence of the field estimate on anatomy \citep{sled.1988}. The next step is the tissue classification into three types: gray matter, white matter and cerebrospinal fluid (CSF). This is critical for identifying the tissue boundaries where surface measurements are obtained. An artificial neural network classifier \citep{kollakian.1996,ozkan.1993}  or Gaussian mixture models  \citep{good.2001} can be used to segment the tissue types automatically. The Statistical Parametric Mapping (SPM) package\footnote{The SPM package is available at {\tt www.fil.ion.ucl.ac.uk/spm}.}  uses a Gaussian mixture with a prior tissue density map.

\begin{figure}
\centering
\includegraphics[width=1\linewidth]{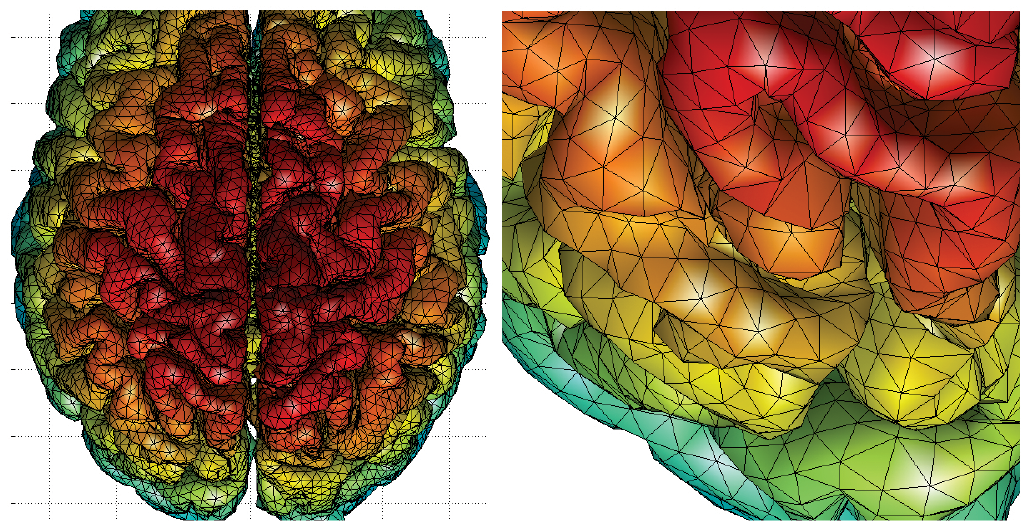}
\caption{\label{fig:cortex1} Left: The outer cortical brain
surface mesh consisting of 81,920 triangles. Measurements are defined at mesh vertices. Right: The part of the mesh is enlarged to show the convoluted nature of the surface.}
\end{figure}

After the segmentation, the tissue boundaries are extracted as triangular meshes. In order to triangulate the boundaries, the marching cubes algorithm \citep{lorensen.1987}, level set method \citep{sethian.2002}, the deformable surfaces method \citep{davatzikos.1995}  or anatomic segmentation using proximities (ASP) method \citep{macdonald.2000} can be used. Brain substructures such as the brain stem and the cerebellum are usually automatically removed in the process. The resulting triangular mesh is expected to be topologically equivalent to a sphere.  For example, the triangular mesh resulted from the ASP method consists of
40,962 vertices and 81,920 triangles with the average internodal
distance of $3 \mbox{ mm}$. Figure \ref{fig:cortex1} shows a representative cortical mesh obtained from ASP. Surface measurements such as cortical thickness can be automatically obtained at each mesh vertex. Subcortical brain surfaces such as amygdala and hippocampus are extracted similarly, but often done in a semi-automatic fashion with the marching cubes algorithm on manual edited subcortical volumes. In the hippocampus case study, the left and right hippocampi were manually segmented in the template using the protocol outlined in \citep{rusch.2001}.

Comparing measurements defined across different cortical surfaces is not trivial due to the fact that no two cortical surfaces are identically shaped. In comparing measurements across different 3D whole brain images, 3D volume-based image registration such as Advanced Normalization Tools (ANTS) \citep{avants.2008} is needed. However, 3D image registration techniques tend to misalign sulcal and gyral folding patterns of the cortex. Hence, 2D surface-based registration is needed in order to  match measurements across different cortical surfaces. Various surface registration methods have been proposed \citep{chung.2005.ipmi,chung.2012.CNA,thompson.1996,davatzikos.1997,miller.1997,fischl.1999}. Most methods solve a complicated optimization  problem of minimizing the measure of discrepancy between two surfaces. A much simpler spherical harmonic representation technique provide a simple way of approximately matching surfaces without time-consuming numerical optimization \citep{chung.2012.CNA}.

Surface registration and the subsequent surface-based analysis usually require parameterizing surfaces. It is natural to assume the surface mesh to be a discrete approximation to the underlying cortical surface, which can be treated as a smooth 2D Riemannian manifold. Cortical surface parameterization has been done previously in \citep{thompson.1996,joshi.1995,chung.2012.CNA}. The surface parameterization also provides surface shape features such as the Gaussian and mean curvatures, which measure anatomical variations associated with the deformation of the cortical surface during, for instance, development and aging \citep{dale.1999,griffin.1994,joshi.1995,chung.2012.CNA}.

\section{Surface Parameterization}
\index{surface ! parameterization}
In order to perform statistical analysis on a surface, parameterization of the surface is often required \citep{chung.2012.CNA}. Brain surfaces are often mapped onto a plane or a sphere.  Then surface measurements defined on mesh vertices are also mapped onto the new domain and analyzed. However, almost all surface parameterizations suffer metric distortions, which in turn influence the spatial covariance structure so it is not necessarily the best approach.

We model the cortical surface $\mathcal{M}$ as a smooth 2D Riemannian manifold parameterized by two parameters $u^1$ and $u^2$ such that any point ${\bf x} \in \mathcal{M}$ can be represented as
$${\bf X}(u^1,u^2) = \{x_1(u^1,u^2),x_2(u^1,u^2),x_3(u^1,u^2):(u^1,u^2) \in D \subset \mathbb{R}^2 \}$$
for some parameter space ${\bf u}=(u^1,u^2) \in D \subset \mathbb{R}^2$ \citep{boothby.1986,carmo.1992,chung.2003.NI,kreyszig.1959}. The aim of the parameterization is estimating the coordinate functions $x_1, x_2, x_3$ as smoothly as possible. 

Both global and local parameterizations are available. A global parameterization, such as tensor B-splines and spherical harmonic representation, are computationally expensive compared to a local surface parameterization. A local surface parameterization in the neighborhood of point ${\bf x}=(x_1, x_2, x_3)$ can be obtained via the projection of the local surface patch onto the tangent
plane $T_{\bf x}(\mathcal{M})$ \citep{chung.2012.CNA,joshi.1995}.

\subsection{Local Parameterization by Quadratic Polynomial}
\index{quadratic polynomial}
\index{surface ! quadratic parameterization}
A local parameterization is usually done by fitting a quadratic polynomial of the form \bqn {\bf X}(u^1,u^2)
=  \beta_1u^1 + \beta_2u^2 + \beta_3(u^1)^2 + \beta_4u^1u^2 +
\beta_5(u^2)^2 \label{eq:quad_poly}\eqn
in $ (u^1, u^2) \in D \subset \mathbb{R}^2$. The data can be centered so there is no constant term in the quadratic form (\ref{eq:quad_poly}) \citep{chung.2012.CNA}. The coefficients $\beta_i$ are usually estimated by the least squares method \citep{joshi.1995,khaneja.1998,chung.2012.CNA}.

In estimating various differential geometric measures such as the Laplace-Beltrami operator or curvatures, it is not necessary to find global surface parameterization of $\mathcal{M}$. Local surface parameterization such as the quadratic polynomial fit is sufficient to obtain such geometric quantities \citep{chung.2012.CNA}. The drawback of the polynomial parameterization is that there is a tendency to weave the outermost mesh vertices to find vertices in the center. Therefore this is not advisable to directly fit (\ref{eq:quad_poly}) when one of the coordinate values rapidly changes.

\subsection{Surface Flattening}

Parameterizing cortical and subcortical surfaces with respect to simpler algebraic surfaces such as a unit sphere is needed to establish a standard coordinate system. However, polynomial regression type of local parameterization is ill-suited for this purpose. For the global surface parameterization, we can use {\em surface flattening} \citep{andrade.2001, angenent.1999}, which is nonparametric in nature. The surface flattening parametrizes a surface by either solving a partial differential equation or optimizing its variational form. 

Deformable surface algorithms naturally provide one-to-one maps from cortical surfaces to a sphere since the algorithm initially starts with a spherical mesh and deform{s} it to match the tissue boundaries \citep{macdonald.2000}. The deformable surface algorithms usually start with the second level of triangular subdivision of {an} icosahedron as the initial surface. After several iterations of deformation and triangular subdivision, the resulting cortical surface contains 
very dense triangle elements. 
There are many  surface flattening techniques such as  conformal mapping  \citep{angenent.1999, gu.2004, hurdal.2004} quasi-isometric mapping \citep{timsari.2000}, area preserving mapping \citep{brechbuhler.1995}, {and the} Laplace equation method \citep{chung.2012.CNA}.

For many surface flattening methods to work, the starting binary object has to be close to star-shape or convex. For shapes with a more complex structure, the methods may  create numerical singularities in mapping to the sphere. Surface flattening  can destroy the inherent geometrical structure of the cortical surface due to the metric distortion. Any structural or functional analysis associated with the cortex can be performed without surface flattening if {the} intrinsic geometric method is used \citep{chung.2012.CNA}.

\subsection{Spherical Harmonic Representation}

\index{representations ! spherical harmonic}
\index{spherical harmonic ! representation}
\index{spherical coordinates}
\index{spherical harmonic}

The spherical harmonic  (SPHARM) representation
\footnote{The SPHARM package is available at\\
{\tt www.stat.wisc.edu/$\sim$mchung/softwares/weighted-SPHARM/weighted-SPHARM.html}} 
is a widely used subcortical surface parameterization technique \citep{chung.2008.sinica,chung.2012.CNA,gerig.2001, gu.2004, kelemen, shen.2004}. SPHARM represents the coordinates of mesh vertices as a linear combination of spherical harmonics. SPHARM has been mainly used as a data reduction technique for compressing global shape features into a small number of coefficients. The main global geometric features are encoded in low degree coefficients while the noise are in high degree spherical harmonics \citep{gu.2004}. The method has been used to model various subcortical structures such as ventricles \citep{gerig.2001}, hippocampi \citep{shen.2004} and cortical surfaces \citep{chung.2008.sinica}. The spherical harmonics have a global support. So the resulting spherical harmonic coefficients contain the global shape features and it is not possible to directly obtain local shape information from the coefficients only. However, it is still possible to obtain local shape information by evaluating the representation at each fixed vertex, which gives the smoothed version of the coordinates of surfaces. In this fashion, SPHARM can be viewed as mesh smoothing \citep{chung.2008.sinica,chung.2014.MICCAI}. In this section, we present a brief introduction of SPHARM within a Hilbert space framework. 

Suppose there is a bijective mapping between the cortical surface $\mathcal{M}$ and a unit sphere $S^2$ obtained through a deformable surface algorithm \citep{chung.2012.CNA}. Consider the parameterization of $S^2$ by
$${\bf X}(\theta, \varphi)=(\sin \theta \cos \varphi, \sin \theta \sin \varphi,
\cos \theta),$$ with $(\theta, \varphi) \in [0,\pi) \otimes [0,2\pi)$. The polar angle $\theta$ is the angle from the north pole and the azimuthal angle $\varphi$ is the angle along the horizontal cross-section. Using the bijective mapping, we can parameterize functional data $f$ with respect to the spherical coordinates
\bqn f(\theta,\varphi) = g(\theta,\varphi) + \epsilon(\theta,\varphi),\label{eq:coordinate.function} \eqn
where $g$ is a unknown smooth coordinate function and $\epsilon$ is a zero mean random field, possibly Gaussian. The error function $\epsilon$ accounts for possible mapping errors. The unknown signal $g$ is then estimated in the finite subspace of $ \mathcal{L}^2(S^2)$, the space of square integrable functions in $S^2$, spanned by spherical harmonics in the least squares fashion \citep{chung.2008.sinica}. 

Previous imaging and shape modeling literature have used the complex-valued spherical harmonics \citep{bulow.2004,gerig.2001,gu.2004,shen.2004}. In practice, however, it is sufficient to use only real-valued spherical harmonics \citep{courant.1953, homeier.1996}, which is more convenient in  setting up a real-valued stochastic model (\ref{eq:coordinate.function}). The relationship between the real- and complex-valued spherical harmonics is given in \citep{blanco.1997,homeier.1996}. The complex-valued spherical harmonics can be transformed into real-valued spherical harmonics using an unitary transform.  
 
The spherical harmonic $Y_{lm}$ of degree $l$ and order $m$ is defined as
\bq
Y_{lm} = 
\begin{cases} 
c_{lm}P^{|m|}_l(\cos\theta)\sin
(|m|\varphi), &-l \leq m\leq -1, \\
 \frac{c_{lm}}{\sqrt{2}}P_l^{|m|}(\cos\theta),& m=0,\\
c_{lm} P^{|m|
}_l(\cos\theta)\cos (|m|\varphi),& 1 \leq m\leq l,
 \end{cases}
\eq
where $c_{lm}=\sqrt{\frac{2l+1}{2\pi}\frac{(l-|m|)!}{(l+|m|)!}}$ and
$P^{m}_l$ is the {\em associated Legendre polynomial} of order $m$ \citep{courant.1953,wahba.1990}, which is given by
\bq P_l^{m}(x) =  \frac{(1-x^2)^{m/2}}{2^l l!}  \frac{d^{l+m}}{dx^{l+m}}
(x^2 -1)^l,   x \in [-1,1].\label{eq:ass.legendre}\eq

The first few terms of the spherical harmonics are
\bq Y_{00}=\frac{1}{\sqrt{4\pi}}, Y_{1,-1} =\sqrt{\frac{3}{4\pi}} \sin \theta \sin \varphi,\\
 Y_{1,0} =  \sqrt{\frac{3}{4\pi}}
\cos \theta, Y_{1,1} = \sqrt{\frac{3}{4\pi}} \sin \theta \cos \varphi.
\eq
The spherical harmonics are orthonormal with respect to the inner product 
\bq \langle f_1, f_2 \rangle &=& \int_{S^2} f_1(\Omega)
f_2(\Omega) \;d\mu(\Omega), \eq where $\Omega=(\theta,\varphi)$ and the Lebesgue measure
$d\mu(\Omega)=\sin \theta d\theta d\varphi$. The norm is then defined as
\bqn  || f_1 ||  = \langle f_1, f_1 \rangle ^{1/2}\label{eq:norm}.\eqn

The unknown mean {function} $g$ is  estimated by minimizing the integral of the squared residual in $\mathcal{H}_k$, the space spanned by up to $k$-th degree spherical harmonics:
\bqn \widehat{g} (\Omega) = \arg\min_{h \in \mathcal{H}_k} \int_{S^2} \Big|f(\Omega)- h(\Omega)\Big|^2\;d\mu(\Omega).  
\label{eq:LSE}\eqn
It can be shown that the minimization is given by
\bqn \widehat{g} (\Omega) = \sum_{l=0}^{k} \sum_{m=-l}^l \langle f, Y_{lm} \rangle Y_{lm}(\Omega), \label{eq:SPHARM}\eqn
the Fourier series expansion. The expansion (\ref{eq:SPHARM}) has been referred to as the {\em spherical harmonic representation} \citep{chung.2008.sinica,gerig.2001,gu.2004,shen.2004,shen.2006}. This technique has been used in representing various brain subcortical structures such as hippocampi  \cite
{shen.2004} and ventricles \citep{gerig.2001} as well as the whole brain cortical surfaces \citep{chung.2008.sinica,gu.2004}. By taking {each component of} Cartesian coordinates of mesh vertices as the functional signal $f$, surface meshes can be parameterized as a function of $\theta$ and $\varphi$.

\index{weighted spherical harmonics }
\index{spherical harmonics}
\index{wavelets}

The spherical harmonic coefficients can be estimated {in least squares fashion}. However, for an extremely large number of vertices and expansions, the least squares method may be difficult to directly invert large matrices. Instead, the {\em iterative residual fitting (IRF)} algorithm \citep{chung.2008.sinica} can be used to iteratively estimate the coefficients by partitioning the larger problem into smaller subproblems. The IRF algorithm is similar to the matching pursuit method \citep{mallat.1993}. The IRF algorithm was developed to avoid the computational burden of inverting a large linear problem while the matching pursuit method was originally developed to compactly decompose a time-frequency signal into a linear combination of a pre-selected pool of basis functions. Although increasing the degree of the representation increases the goodness-of-fit, it also increases the number of estimated coefficients quadratically. So it is necessary to stop the iteration at the specific degree $k$, where the goodness-of-fit and the number of coefficients balance out. This idea was used in determining the optimal degree of SPHARM \citep{chung.2008.sinica}.

The limitation of SPHARM is that it produces the Gibbs phenomenon, i.e., ringing artifacts,  for discontinuous and rapidly changing continuous measurements \citep{chung.2008.sinica,gelb.1997}. The Gibbs phenomenon can be effectively removed by weighting the spherical harmonic coefficients exponentially smaller, which {makes the representation smooth out} rapidly changing signals. The weighted version of  SPHARM is related to isotropic diffusion smoothing \citep{andrade.2001,cachia.tmi.2003,chung.2012.CNA,chung.2005.ipmi} as well as the diffusion wavelet transform \citep{chung.2008.sinica, chung.2014.MICCAI,hosseinbor.2014.miccai}.

\section{Surface Registration}
\index{surface ! registration}

To construct a test statistic locally at each vertex across different surfaces, {one must register} the surfaces to a common template surface. Nonlinear cortical surface registration is often performed by minimizing objective functions that measure the global fit of two surfaces while maximizing the smoothness of the deformation in such a way that the cortical gyral or sulcal patterns are matched smoothly \citep{chung.2005.ipmi, robbins.2003, thompson.1996}. There is also a much simpler way of aligning surfaces using SPHARM representation \citep{chung.2008.sinica}. Before any sort of nonlinear registration is performed, an affine registration is performed to align and orient the global brain shapes. 

\subsection{Affine Registration}

Anatomical objects extracted from 3D medical images are aligned using affine transformations to remove the global size differences. Affine registration requires identifying corresponding landmarks either manually or automatically. The affine transform $T$ of point $p=(p_1, \cdots, p_d)' \in \mathbb{R}^d$ to $q=(q_1, \cdots, q_d)'$ is given by
$$q = Rp + c,$$
where the matrix $R$ corresponds to rotation, scaling and shear while $c$ corresponds to translation. 
Although the affine transform is not linear, it can be  made into a linear form by augmenting the transform. The affine transform can be rewritten  as
\bqn
\left(
\begin{array}{c}
 q   \\
1   \\
\end{array}
\right)
=
\left(
\begin{array}{cc}
 R &  c    \\
 0 \cdots 0&   1   \\
\end{array}
\right) \left(
\begin{array}{c}
 p   \\
1   \\
\end{array}
\right).
\label{affine-matrixform}
\eqn

Let $$A= \left(
\begin{array}{cc}
 R &  c    \\
 0 \cdots 0&   1   \\
\end{array}
\right).$$
{Trivially,} $A$ is linear a linear operator. The matrix $A$ is the most often used form for recording the affine registration.

Let ${\bf p}_i$ be the $i$-th landmark  and its corresponding affine transformed points ${\bf q}_i$. Then we rewrite (\ref{affine-matrixform}) as 
\bq
\underbrace{\left(
\begin{array}{ccc}
{\bf q}_1 & \cdots &{\bf q}_n 
\end{array}
\right)}_{Q}
=
\left(
\begin{array}{cc}
 R &  c    \\
\end{array}
\right) 
\underbrace{
\left(
\begin{array}{ccc}
{\bf  p}_1& \cdots & {\bf p}_n\\
 1 &\cdots & 1
\end{array}
\right)}_{P}.
\eq
Then the least squares estimation is given as
$$\left(
\begin{array}{cc}
 \widehat{R} & \widehat{ c}    \\
\end{array}
\right) 
= QP' (PP')^{-1}.
$$
Then the points ${\bf p}_i$ are mapped to $\widehat{R}{\bf p}_i + \widehat{c}$, which may not coincide with ${\bf q}_i$ in general.  
In practice, landmarks are automatically identified from T1-weighted MRI.

\subsection{SPHARM Correspondence}
Using SPHARM, it is possible to approximately register surfaces with different mesh topology without any  optimization. The crude alignment can be done by coinciding the first order ellipsoid meridian and equator in the SPHARM representation \citep{gerig.2001, styner.2006}. However, this can be improved. Consider SPHARM representation of surface $H = (h_1, h_2, h_3)$ with spherical angles $\Omega$ given by
$$h_i(\Omega) = \sum_{l=0}^k \sum_{m=-l}^l h_{lm}^i Y_{lm}(\Omega),$$
where $(v_1, v_2, v_3)$ are the coordinates of mesh vertices and SPHARM coefficient $ h_{lm}^i =\langle v_i, Y_{lm}\rangle$. 
Consider another SPHARM representation $J=(j_1, j_2, j_3)$ obtained from mesh coordinate $w_i$:
$$j_i(\Omega) = \sum_{l=0}^k \sum_{m=-l}^l w_{lm}^i  Y_{lm}(\Omega),$$
where $w_{lm}^i  = \langle w_i, Y_{lm}  \rangle$. Suppose the surface $h_i$ is deformed to $h_i + d_i$ by the amount of  displacement $d_i$. We wish to find $d_i$ that minimizes the discrepancy between $h_i +d_i$ and $j_i$ in the subspace $\mathcal{H}_k$ spanned using up to the $k$-th degree spherical harmonics. It can be shown that 
\bqn  \arg \min_{d_i \in \mathcal{H}_k} \Big|\Big| \widehat{h_i} + d_i -\widehat{j_i} \Big|\Big| 
= \sum_{l=0}^k \sum_{m=-l}^l (w_{lm}^i-v_{lm}^i)Y_{lm}(\Omega).\eqn
This implies that the optimal displacement of matching two surfaces is obtained by simply taking the difference between two SPHARM and matching coefficients of the same degree and order. Then a specific point $\widehat{h_i}(\Omega)$ in one surface corresponds to $\widehat{j_i}(\Omega)$ in the other surface. We refer to this point-to-point surface matching as the {\em SPHARM correspondence} \citep{chung.2008.sinica}. Unlike other surface
registration methods \citep{chung.2005.ipmi, robbins.2003, thompson.1996}, it is not necessary to consider an additional cost function that guarantees the smoothness of the displacement
field since the displacement field $d=(d_1, d_2, d_3)$ is already a
linear combination of smooth basis functions. 

\begin{figure}
\centering
\includegraphics[width=1\linewidth]{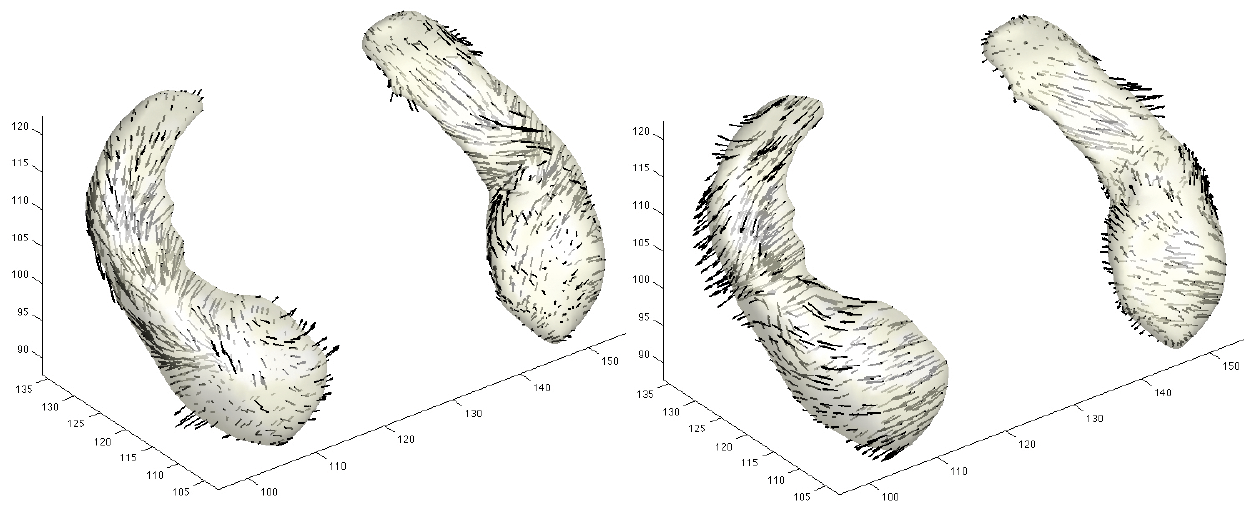}
\caption{\label{fig:arrows} The displacement vector fields of registering from the hippocampus template to two individual surfaces. The displacement vector field is obtained from the diffeomorphic image registration \citep{avants.2008}. The variability in the displacement vector field can be analyzed using a multivariate general linear model \citep{chung.2012.CNA}.}
\end{figure}

\subsection{Diffeomorphic Registration}
Diffeomorphic image registration is a recently popular technique for registering volume and surface data \citep{avants.2008,miller.2009,qiu.2006.NI,qiu.2008,yang.2011}. From the affine transformed individual surfaces $\mathcal{M}_j$, an additional nonlinear surface registration to the template using the large deformation diffeomorphic metric mapping (LDDMM) framework can be performed \citep{miller.2009, qiu.2006.NI,qiu.2008,yang.2011}. In LDDMM, given the template surface $\mathcal{M}$, the diffeomorphic transformations, which are one-to-one, smooth forward, and inverse transformation, are constructed as follows. We estimate the diffeomorphism $d$ between them as a smooth streamline given by the Lagrangian evolution:
$$\frac{\partial d}{\partial t} (x,t) = v \circ d(x,t)$$
with $d(x,0) = x, \; t \in [0,1]$ for time-dependent velocity field $v$. 
Note the surfaces $\mathcal{M}_j$ and $\mathcal{M}$ are the start and end points of the diffeomorphism, i.e. $\mathcal{M}_j \circ d(\cdot, 0) =  \mathcal{M}_j$ and $\mathcal{M}_j \circ d(\cdot, 1) =  \mathcal{M}$.
By solving the evolution equation numerically, we obtain the diffeomorphism. By averaging the inverse deformation fields from the template $\mathcal{M}$ to individual subjects, we may obtain yet another final more refined template. The vector fields $v$ are constrained to be sufficiently smooth {to generate} diffeomorphic transformations over finite time \citep{dupuis.1998}. Figure \ref{fig:arrows} shows the resulting displacement vector fields of warping from the template to two hippocampal surfaces. In the deformation-based morphometry, variability in the displacement is {used  to characterize} surface growth and differences \citep{chung.2012.CNA}.

\section{Cortical Surface Features}
The human cerebral cortex has the topology of a 2D highly convoluted grey matter shell with an average thickness of {3 mm} \citep{chung.2012.CNA,macdonald.2000}. The outer boundary of the shell is called  the {\em outer cortical surface} while the inner boundary is called the {\em inner cortical surface}. Various cortical surface features such as curvatures, local surface area and cortical thickness have been used in {quantifying anatomical variations}. Among them, cortical thickness has been more often analyzed than other features. 

\subsection{Cortical Thickness}
Once we extract both the outer and inner cortical surface meshes, {\em cortical thickness} can be computed at each mesh vertex. The cortical thickness is defined as  the distance between the corresponding vertices between the inner and outer surfaces \citep{macdonald.2000}. There are many different computational approaches in measuring cortical thickness. 
In one approach, the vertices on the inner triangular mesh are  deformed to fit the outer surface by minimizing a cost function that involves bending, stretch and other topological
constraints \citep{chung.2012.CNA}. There is also an alternate method for automatically measuring cortical thickness based on the Laplace equation \citep{jones.2000}.

The average cortical thickness for each individual is about {3 mm} \citep{henery.1989}. Cortical thickness varies from
1 to {4 mm} depending on the location of the cortex. In normal
brain development, it is highly likely that the change of cortical
thickness may not be uniform across the cortex. Since different clinical populations are expected to show different patterns of cortical thickness variations, cortical thickness {has also been} used as a quantitative index for characterizing a clinical population \citep{chung.2005.ipmi}. Cortical thickness varies locally by region and is likely to be influenced by aging, development and disease \citep{barta.2005}. By analyzing how cortical thickness varies between clinical and non-clinical populations, we can locate regions on the brain related to a specific pathology.

\subsection{Surface Area and Curvatures}
\index{local surface area}

As in the case of local volume changes in the deformation-based morphometry, the rate of cortical surface area
expansion or reduction may not be uniform across the cortical
surface \citep{chung.2012.CNA}. Suppose that cortical surface $\mathcal{M}$ is parameterized by the parameters ${\bf u}=(u^1,u^2)$ such
that any point ${\bf x} \in \mathcal{M}$ can be written as
${\bf x} = {\bf X}({\bf u})$. Let ${\bf X}_i=\partial {\bf
X}\slash \partial u^i$ be the partial derivative vectors. The {\em
Riemannian metric tensor} $g_{ij}$ is then given by the inner product
between two vectors ${\bf X}_i$ and ${\bf X}_j$, i.e. 
$$g_{ij}(t) = \langle {\bf X}_i, {\bf X}_j \rangle.$$ 
The tensor $g_{ij}$ measures the amount of the deviation of the
cortical surface from a flat Euclidean plane and can be used to measure lengths, angles and areas on the cortical surface.

Let $g =(g_{ij})$ be a $2 \times 2$ matrix
of metric tensors. The total surface area of the cortex $\mathcal{M}$ is then given
by $$\int_D \sqrt{\det g} \;d{\bf u},$$ where $D=X^{-1}(\mathcal{M})$ is
the parameter space \citep{kreyszig.1959}. The term $\sqrt{\det g}$ is
called the {\em  surface area element} and it
measures the transformed area of the unit square in the parameter
space $D$ via transformation $X: D \to \mathcal{M}$. The surface area element can be
considered as the generalization of the Jacobian determinant, which is  used in
measuring local volume changes in the tensor-based morphometry \citep{chung.2012.CNA}. 

Instead of using the metric tensor formulation, it is possible to
quantify local surface area change in terms of the areas of the
corresponding triangles in surface meshes. However, this formulation assigns the computed surface area to each face instead of each vertex. This causes a problem in both surface-based smoothing and statistical analysis, where values are required to be given on vertices.
Interpolating scalar values on vertices from face values can be done by
the weighted average of face values. It is not hard to develop surface-based
smoothing and statistical analysis on  face values, as a form of dual formulation, but
the cortical thickness and the curvature metric are defined on
vertices so we end up with two separate approaches: one for
metrics defined on vertices and the other for metrics defined on
faces. Therefore, the metric tensor approach provides a better
unified quantitative framework for the subsequent statistical analysis. 

\index{curvatures}
\index{surface ! curvatures}

The {\em principal curvatures} characterize the shape and location of
the sulci and gyri, which are the valleys and crests of the
cortical surfaces \citep{bartesaghi.2001,joshi.1995,khaneja.1998}. By measuring the curvature changes, rapidly folding and cortical
regions can be localized. The principal curvatures $\kappa_1$ and $\kappa_2$ can be represented as functions of $\beta_i$ in quadratic surface (\ref{eq:quad_poly}) \citep{boothby.1986, kreyszig.1959}.

\subsection{Gray Matter Volume}
\index{local gray matter volume}
\index{gray matter}

Local volume can be computed using the determinant of the Jacobian
of deformation and used in detecting the regions of brain
tissue growth and loss {in brain development}  \citep{chung.2012.CNA}. Compared to the local surface area change, the local volume change measurement is more sensitive to small deformation of the brain. If a unit cube
increases its sides by one, the surface area will increase by
$2^2-1=3$ while the volume will increase by $2^3-1=7$. Therefore,
the statistical analysis based on the local volume change is
at least twice more sensitive compared to that of the local
surface area change. So the local volume change should be able to
pick out gray matter {tissue growth patterns} even when the local
surface area change may not. 

The gray matter can be considered as a thin shell
bounded by two surfaces with varying cortical thickness.  In most deformable surface algorithms like FreeSurfer, each triangle on the outer surface has a corresponding triangle on the inner
surface. Let ${\bf p}_1,{\bf p}_2,{\bf
p}_3$ be the three vertices of a triangle on the outer surface and
${\bf q}_1,{\bf q}_2,{\bf q}_3$ be the corresponding three
vertices on the inner surface such that ${\bf p}_i$ is linked to
${\bf q}_i$. 
Then the volume of the triangular prism is given by the sum of the determinants
$$D({\bf p}_1,{\bf p}_2,{\bf p}_3,{\bf q}_1)
+D({\bf p}_2,{\bf p}_3,{\bf q}_1,{\bf q}_2)+ D({\bf p}_3,{\bf
q}_1,{\bf q}_2,{\bf q}_3)$$ where 
$$D({\bf a},{\bf b},{\bf c},{\bf
d})=|\det({\bf a}-{\bf d},{\bf b}-{\bf d}, {\bf c}-{\bf d})|/6$$ is
the volume of a tetrahedron whose vertices are $\{{\bf a},{\bf
b},{\bf c},{\bf d}\}$. Afterwards, the total gray matter volume
 can be estimated by summing the volumes of all
triangular prisms \citep{chung.2012.CNA}.

\section{Surface Data Smoothing}

Cortical surface mesh extraction and cortical thickness computation are expected to introduce noise \citep{chung.2005.ipmi, fischl.2000, macdonald.2000}. To counteract this, surface-based data smoothing is necessary \citep{andrade.2001, cachia.mia.2003, cachia.tmi.2003, chung.2012.CNA}. For 3D whole brain volume data,  Gaussian kernel smoothing is desirable in many statistical analyses \citep{dougherty.1999,rosenfeld.1982}. Gaussian kernel smoothing weights neighboring observations according to their 3D Euclidean distance. Specifically, Gaussian kernel smoothing of functional data or image $f({\bf x}), {\bf
x}=(x_1,\dots,x_n) \in {\mathbb{R}}^n$ with {\em full width
at half maximum (FWHM}) $=4(\ln 2)^{1/2}\sqrt t$ is defined as the
convolution of the Gaussian kernel with $f$: 
\bqn F({\bf x},t)=
\frac{1}{(4\pi t)^{n/2}}\int_{{\mathbb{R}}^n} e^{-(x-y)^2/4t} f(y)
dy. \label{eq:convolution}\eqn 
However, due to the convoluted nature of the cortex, whose geometry is non-Euclidean, we {cannot} directly
use the formulation (\ref{eq:convolution}) on the cortical surface. For data that lie on a 2D surface, smoothing must be weighted according to the geodesic distance along the surface, which is not straightforward \citep{andrade.2001,chung.2012.CNA,chung.2001.diffusion}.

\subsection{Diffusion Smoothing}
\label{sec:diffusionsmootihng}

\index{smoothing ! diffusion}
\index{diffusion smoothing}
\index{surface-based diffusion smoothing}

By formulating Gaussian kernel smoothing as a solution of a diffusion equation on a Riemannian manifold, the Gaussian kernel smoothing approach can be generalized to an arbitrary curved
surface. This generalization is called {\em diffusion smoothing} and {was first introduced} in the analysis of fMRI data on the cortical surface \citep{andrade.2001} and cortical thickness \citep{chung.2001.diffusion} in 2001.

It can be shown that Gaussian kernel smoothing
(\ref{eq:convolution}) is the integral solution of the
$n$-dimensional diffusion equation 
\bqn 
\frac{\partial F}{\partial t}  = \Delta F
\eqn 
with the initial condition $F({\bf
x},0)=f({\bf x})$, where 
$$\Delta=
\frac{\partial^2}{\partial x_1^2} + \cdots + \frac{\partial^2}{\partial x_n^2}$$ is the Laplacian in
$n$-dimensional Euclidean space. Hence the Gaussian kernel
smoothing is equivalent to the diffusion of the initial data $f({\bf x})$ after time $t$.

Diffusion equations have been widely used in image processing as a form of noise reduction starting with Perona and Malik in 1990 \citep{perona.1990}. Numerous diffusion techniques have been developed for surface data smoothing  \citep{andrade.2001,chung.2012.CNA,cachia.tmi.2003, cachia.mia.2003, chung.2005.ipmi, joshi.2009}. 
When applying diffusion smoothing on curved surfaces, the smoothing somehow has to incorporate the geometrical features of the curved surface and the Laplacian $\Delta$ should change accordingly. The extension of the Euclidean Laplacian to an arbitrary Riemannian manifold is called the {\em Laplace-Beltrami operator} \citep{arfken.2000, kreyszig.1959}. 
The approach taken in \citep{andrade.2001}  is based on a local flattening of the cortical surface and estimating the planar Laplacian, which may not be as accurate as the cotan estimation based on the finite element method (FEM) given in \citep{chung.2001.diffusion}. Further, {the FEM approach} completely avoids the use of surface flattening and parameterization; thus, it is more robust.

For given Riemannian metric tensor $g_{ij}$, the Laplace-Beltrami operator $\Delta$ is defined as
\bqn \Delta F = \sum_{i,j}
\frac{1}{|g|^{1/2}}  \frac{\partial }{\partial u^i} \Big(|g|^{1/2}g^{ij} \frac{\partial F}{\partial u^j} \Big) 
\label{eq:Laplace-Beltrami},
\eqn where $(g^{ij})=g^{-1}$ \citep{arfken.2000}. Note that when $g$ becomes an identity matrix, the Laplace-Beltrami operator reduces to the standard 2D Laplacian:
$$\Delta F= \frac{\partial^2 F}{\partial (u^1)^2} +
\frac{\partial^2 F}{\partial (u^2)^2}.$$
Using FEM on the triangular cortical mesh, it is possible to estimate the Laplace-Beltrami
operator as the linear weights of neighboring vertices using the cotan formulation, which is first given in \citep{chung.2001.diffusion}.

\begin{figure}
\centering
\includegraphics[width=0.6\linewidth]{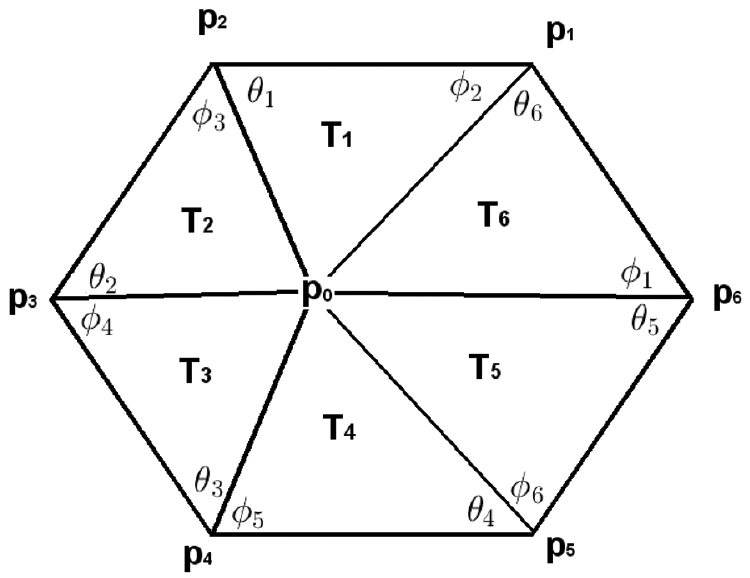}
\caption{\label{fig:hexagon1} A typical triangulation in the
neighborhood of ${\bf p} ={\bf p}_0$ in a surface mesh.}
\end{figure}

Let ${\bf p}_1,\cdots,{\bf p}_m$ be $m$ neighboring vertices around
the central vertex ${\bf p}={\bf p}_0$ (Figure \ref{fig:hexagon1}). Then the estimated
Laplace-Beltrami operator is given by
\bqn \widehat {\Delta F}({\bf p}) = \sum_{i=1}^m
w_i\big(F({\bf p}_i)-F({\bf p})\big) \label{eq:FEMLB}\eqn
with the weights
$$w_i= \frac{\cot \theta_i + \cot \phi_i}{\sum_{i=1}^m \|T_i\|},$$
where  $\theta_i$ and $\phi_i$ are the two angles opposite to
the edge connecting ${\bf p}_i$ and ${\bf p}$, and $\|T_i\|$
is the area of the $i$-th triangle (Figure \ref{fig:hexagon1}).

FEM estimation (\ref{eq:FEMLB}) is an improved formulation from the previous attempt in
diffusion smoothing \citep{andrade.2001}, where the Laplacian is simply estimated as the planar
Laplacian after locally flattening the triangular mesh consisting
of nodes ${\bf p}_0,\cdots,{\bf p}_m$ onto a flat plane. 
Afterwards, the finite difference (FD) scheme can be used to iteratively solve the diffusion equation at each vertex ${\bf p}$:
$$\frac{F({\bf p},t_{n+1}) - F({\bf p},t_n)}{t_{n+1}-t_n} = \widehat {\Delta} F({\bf p},t_n),$$ with the initial condition $F({\bf p},0)=f({\bf p})$ and fixed $\delta t = t_{n+1} - t_n$. After $N$ iterations, the FD gives the diffusion of the initial data $f$ after time $N\delta t$. If the diffusion
were applied to Euclidean space, it would be approximately equivalent to
Gaussian kernel smoothing with
$$ \mbox{FWHM} =4(\ln2)^{1/2}\sqrt{N\delta t}.$$
For large meshes, computing the linear weights for the Laplace-Beltrami operator
takes a fair amount of time, but once the weights are computed, it is
applied repeatedly throughout the iterations as a matrix multiplication. Unlike Gaussian kernel smoothing, diffusion smoothing
is an iterative procedure.

\subsection{Iterated Kernel Smoothing}
\index{iterated kernel smoothing}
\index{smoothing ! iterated}

Diffusion smoothing  use FEM and FD, {which are known} to suffer numerical instability if sufficiently small step size is not chosen in the forward Euler scheme. To remedy the problem associated with diffusion smoothing, {\em iterative kernel smoothing}\footnote{MATLAB package: {\tt www.stat.wisc.edu/$\sim$mchung/softwares/hk/hk.html}.}  was introduced \citep{chung.2005.ipmi}. The method has been used in smoothing various cortical surface data: cortical curvatures \citep{luders.ni.2006,gaser.2006}, cortical thickness \citep{luders.hbm.2006,bernal.2008}, hippocampus \citep{shen.CSB.2006,zhu.2007}, magnetoencephalography (MEG) \citep{han.2007} and functional-MRI \citep{hagler.2006,jo.2007}. This and its variations are probably the most widely used method for smoothing brain surface data at this moment. In iterated kernel smoothing, kernel weights are spatially adapted to follow the shape of the heat kernel in a discrete fashion.

The {\em $n$-th iterated kernel smoothing} of signal $f \in L^2(\mathcal{M})$ with kernel $K_{\sigma}$ is defined as 
$$K_{\sigma}^{(n)} * f (p) =\underbrace{K_{\sigma}
*\cdots * K_{\sigma}}_{n  \mbox{ \small{times} }}*f (p),$$
where $\sigma$ is the bandwidth of the kernel. 
If $K_{\sigma}$ is a heat kernel, we have the following iterative relation \citep{chung.2005.ipmi}:
\bqn 
K_{\sigma} * f (p)= K_{\sigma/n}^{(n)} * f (p).
\label{eq:heatiteration}
\eqn
The relation (\ref{eq:heatiteration}) shows  that kernel smoothing with large bandwidth $\sigma$ can be decomposed into $n$ repeated applications of kernel smoothing with smaller bandwidth $\sigma/n$. This idea can be used to approximate the heat kernel. When the bandwidth is small, the heat kernel behaves like the Dirac-delta function and, using the parametrix expansion \citep{rosenberg.1997, wang.1997}, we can approximate it locally using the Gaussian kernel.

Let ${\bf p}_1,\cdots, {\bf p}_m$ be $m$ neighboring vertices of vertex ${\bf p}={\bf p}_0$ in the mesh. The geodesic distance $d({\bf p}, {\bf p}')$ between ${\bf p}$ and its adjacent vertex ${\bf p}_i$ is the length of edge between these two vertices in the mesh. 
Then the discretized and normalized  heat kernel is given by
$$W_{\sigma}({\bf p},{\bf p}_i)=\frac{\exp\big(-\frac{d({\bf p}, {\bf p}_i)^2}
{4 \sigma}\big)}{\sum_{j=0}^m\exp\big(-\frac{d({\bf p} -{\bf p}_j)^2}
{4\sigma}\big)}.$$
Note $\sum_{i=0}^m W_{\sigma}({\bf p},{\bf p}_i) = 1$. For small bandwidth, all the kernel weights are concentrated near the center, so we only need to worry about the first neighbors of a given vertex in a surface mesh. The discrete version of heat kernel smoothing on a triangle mesh is then given by
$$W_{\sigma}*f(p)=\sum_{i=0}^m W_{\sigma}(p,p_i)f(p_i)
\label{eq:kern2}.$$ 
The discrete kernel smoothing should converge to heat kernel smoothing as the mesh resolution increases. This is the form of the Nadaraya-Watson estimator \citep{chaudhuri.2000} in statistics. Instead of performing a single kernel smoothing with large bandwidth $n\sigma$, we perform $n$ iterated kernel smoothing with small  bandwidth $\sigma$ as follows $W_{\sigma}^{(n)}*f(p)$. 

\subsection{Heat Kernel Smoothing}
\index{heat kernel smoothing}
\index{smoothing ! heat kernel}

The recently proposed {\em heat kernel smoothing}\footnote{MATLAB package:  {\tt brainimaging.waisman.wisc.edu/$\sim$chung/lb}.} framework constructs the heat kernel analytically using the eigenfunctions of the Laplace-Beltrami operator \citep{seo.2010.miccai}. This method avoids the need for the linear approximation used in iterative kernel smoothing that compounds the approximation error at each iteration. The method represents isotropic heat diffusion analytically as a series expansion so it avoids the numerical convergence issues associated with solving the diffusion equations numerically \citep{andrade.2001,chung.2012.CNA,joshi.2009}. This technique is different from other  diffusion-based smoothing methods in that it bypasses the various numerical problems {such as numerical instability}, slow convergence, and accumulated linearization error. 

Although recently there have been a few studies that introduce heat kernel in computer vision and machine learning \citep{belkin.2006}, they mainly use heat kernel to compute shape descriptors  \citep{sun.2009,bronstein.2010} or to define a multi-scale metric \citep{deGoes.2008}. These studies did not use heat kernel in smoothing functional data on manifolds. 
Further, most kernel methods in machine learning deal with the linear combination of kernels as a solution to penalized regressions, which significantly differs from the heat kernel smoothing framework, which does not have a penalized cost function. There are log-Euclidean and exponential map frameworks on manifolds, where the main interest is in computing the Fr\'{e}chet mean along the tangent space \citep{davis.2010,fletcher.2004}. Such approaches or related methods in \citep{gerber.2009}, the Nadaya-Watson type of kernel regression is reformulated to learn the shape or image means in a population. In the heat kernel smoothing framework, we are not dealing with manifold data but {scalar data} defined on a manifold, so there is no need for exploiting the manifold structure of the data itself. 

\begin{figure}
\centering
\includegraphics[width=1\linewidth]{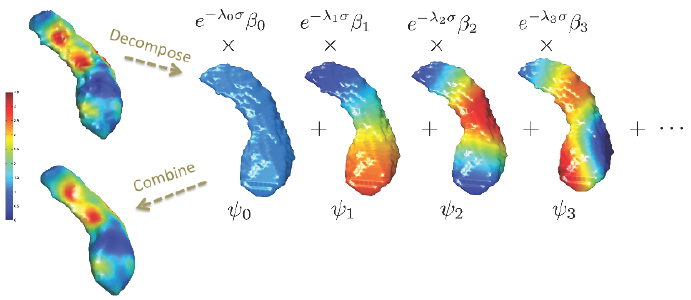} 
 \caption{Schematic of hat kernel smoothing on a hippocampal surface. Given a noisy functional data on the surface, the Laplace-Beltrami eigenfunctions $\psi_j$ are computed and their exponentially weighted Fourier coefficients $\exp{-\lambda_j \sigma}$ are multiplied as a form a regression. This process {smoothes} out the noisy functional signal with bandwidth $\sigma$.}
\label{fig:hippo-basis}
\end{figure}

Let $\Delta$ be the Laplace-Beltrami operator on $\mathcal{M}$. Solving the eigenvalue equation
\bqn 
\Delta \psi_j = -\lambda \psi_j \label{eq:eigensystem},
\eqn 
we order eigenvalues $$0 =  \lambda_0 < \lambda_1 \leq \lambda_2 \leq \cdots, $$ and corresponding eigenfunctions $\psi_0, \psi_1, \psi_2, \cdots$ \citep{rosenberg.1997,chung.2005.ipmi,levy.2006,shi.2009}. The eigenfunctions $\psi_j$ form an orthonormal basis in $L^2(\mathcal{M})$, the space of square integrable functions in $\mathcal{M}$.  Figure \ref{fig:hippo-basis} shows the first four LB-eigenfunctions on a hippocampal surface. 

There is extensive literature on the use of eigenvalues and eigenfunctions of the Laplace-Beltrami operator in medical imaging and computer vision \citep{levy.2006,qiu.2006,reuter.2009.CAD,reuter.2010.IJCV,zhang.2007,zhang.2010}. The eigenvalues have been used in caudate shape discriminators \citep{niethammer.2007}. Qiu et al. used eigenfunctions in constructing splines on cortical surfaces \citep{qiu.2006}. Reuter used the topological features of eigenfunctions \citep{reuter.2010.IJCV}. Shi et al. used the Reeb graph of the second eigenfunction in shape characterization and landmark detection in cortical and subcortical structures \citep{shi.2008.miccai,shi.2009}. Lai et al. used the critical points of the second eigenfunction as anatomical landmarks for colon surfaces \citep{lai.2010}.

Using the eigenfunctions, {\em heat kernel}  $K _{\sigma}(p,q)$ is  
defined as 
\begin{equation}\label{eq:hk}
	K _{\sigma}(p,q) = \sum_{j=0}^{\infty} e^{-\lambda_j \sigma} \psi_j(p) \psi_j (q),
\end{equation}
where $\sigma$ is the bandwidth of the kernel. The heat kernel is the generalization of a Gaussian kernel.
Then {\em heat kernel smoothing} of functional measurement $Y$ is defined 
as
\begin{equation}\label{eq:al_smt}
K_{\sigma} \ast Y(p) = \sum_{j=0}^{\infty} e^{-\lambda_j \sigma} \beta_j \psi_j (p),
\end{equation}
where $\beta_j = \langle Y,\psi_j \rangle$ are Fourier coefficients \citep{
chung.2005.ipmi}. Kernel smoothing $K_{\sigma}*Y$ is taken as the estimate for the unknown mean signal $\theta$. 
The degree for truncating the series expansion can be automatically determined using the forward model selection procedure \citep{chung.2008.sinica}. Figure \ref{fig:hippo-basis}  shows the heat kernel smoothing results with the bandwidth $\sigma=0.5$ and $k=500$ number of LB-eigenfunctions.

Unlike previous approaches to heat diffusion \citep{andrade.2001,chung.2012.CNA,joshi.2009,tasdizen.osher.2006}, heat kernel smoothing avoids the direct numerical discretization of the diffusion equation. Instead, we discretize the basis functions of given manifold $\mathcal{M}$ by solving for the eigensystem (\ref{eq:eigensystem}) and obtain $\lambda_j$ and $\psi_j$. This provides more robust stable smoothing results compared to diffusion smoothing or iterated kernel smoothing approaches.

\section{Statistical Inference on Surfaces}

Surface measurements such as cortical thickness, curvatures, or fMRI responses can be
modeled as random fields on the cortical surface:
\bqn Y(x) = \mu(x) + \epsilon(x), x \in \mathcal{M}, \label{eq:surface-error}\eqn where the
deterministic part $\mu$ is the unknown mean  of the observed functional measurement $Y$
and $\epsilon$ is a mean zero random field. The functional measurements on the brain surface is often modeled using the general linear models (GLMs) or its multivariate version. Various statistical models are proposed for estimating and modeling the signal component $\mu(x)$ \citep{joshi.1995,chung.2012.CNA,chung.2008.sinica} but the majority of the methods are all based on GLM. GLMs have been implemented in the brain image analysis packages such as SPM and fMRISTAT\footnote{The fMRISTAT package is available at {\tt www.math.mcgill.ca/keith/fmristat}.}. 

\subsection{General Linear Models}
\label{sec:GLM}
\index{general linear model}

We set up a GLM at each mesh vertex. Let $y_i$ be the response variable, which is mainly coming from images and ${\bf x}_i=(x_{i1},\cdots, x_{ip})$ to be the variables of interest and ${\bf
z}_i=(z_{i1}, \cdots, z_{ik})$ to be nuisance variables corresponding to the $i$-th subject. Assume there are $n$ subjects, i.e., $i=1, \cdots, n$. We are interested in testing the significance of variables ${\bf x}_i$ while accounting for nuisance covariates ${\bf z}_i$. Then we set up GLM 
$$ y_i = {\bf z}_i{\boldsymbol \lambda}+ {\bf x}_i {\boldsymbol \beta}+\epsilon_i$$
where ${\boldsymbol \lambda} = (\lambda_1,\cdots,\lambda_k)'$ and ${\boldsymbol \beta} =
(\beta_1,\cdots,\beta_p)'$ are unknown parameter vectors to be estimated. 
We assume $\epsilon$ to be the usual zero mean Gaussian noise{, although} the distributional assumption is not required for the least squares estimation. We test hypotheses
$$H_0: {\boldsymbol \beta} =0 \mbox{  vs. } H_1: {\boldsymbol \beta} \neq 0.$$
Subsequently the inference is done by constructing the $F$-statistic with $p$ and $n-p-k$ degrees of freedom. GLMs have been used in quantifying cortical thickness, for instance, in child development \citep{chung.2012.CNA,chung.2005.ipmi} and amygdala shape differences in autism \citep{chung.2012.CNA}. 

In the hippocampus case study, the first T1-weighted MRI scans are taken at $11.6 \pm 3.7$ years for  $n= 124$ children using a 3T GE SIGNA scanner. Variables {\tt age} and {\tt gender} are available.  We also have variable {\tt income}, which is a binary dummy variable indicating whether the subjects are from high- or low-income families. {A total of} 124 children and adolescents are from high- ($> 75000\$ $; $n=86$) and low-income ($<35000 \$ $, $n=38$) parents respectively. In addition to this cross-sectional data, longitudinal data was available for 82 of these subjects ($n=66$, $> 75000\$ $; $n=16$, $< 35000 \$$). The second MRI scan was acquired for these 82 subjects about 2 years later at $14 \pm 3.9$ years. For now, we will simply ignore the correlation between the scans within a subject and will treat them as independent. 

\begin{figure}
\centering
\includegraphics[width=1\linewidth]{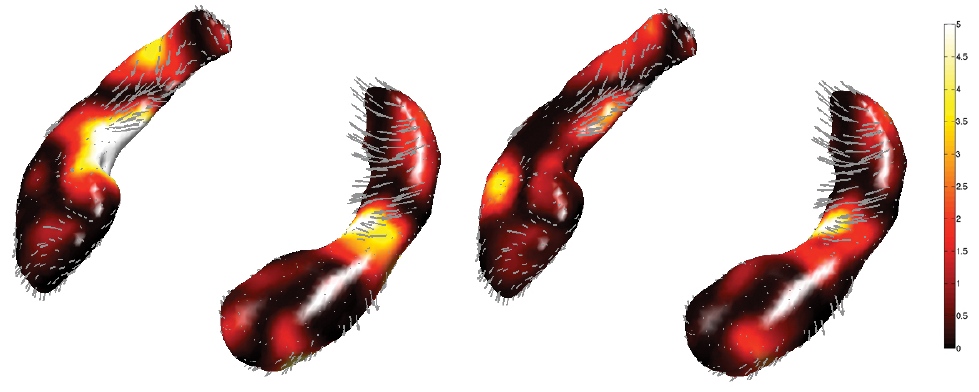} 
 \caption{$F$-statistics maps on testing the {significant} hippocampus shape difference on  the income level while controlling for age and gender. The arrows show the deformation differences  between the groups (high income $-$ low income). The fixed-effects result (left) is obtained by treating the repeat scans as independent. The mixed-effects result (right) is obtained by explicating modeling the covariance structure of the repeat scans with a subject. Both results are not statistically significant under multiple comparisons even at 0.1 level.}
\label{fig:hippo-fstat}
\end{figure}

On the template surface, we have the displacement vector fields of mapping from the template to individual subjects (Figure \ref{fig:arrows}). We take the length of the surface displacement, denoted as {\tt deformation}, with respect to the template as the response variable. The displacement  measures the  shape difference with respect to the template. However, since the length measurement is noisy, surface-based smoothing is necessary. We have used heat kernel smoothing to smooth out noise with the bandwidth 1 and 500 LB-eigenfunctions. Then we set up the GLM:
$$ {\tt deformation} = \lambda_1 + \lambda_2  {\tt age}+ \beta_1 {\tt income} + \epsilon \label{eq:surface-fixed}$$
and test for the significance of $\beta_1$ at each mesh vertex. Figure \ref{fig:hippo-fstat}-left shows the $F$-statistic result on testing $\beta_1$. 

\subsection{Multivariate General Linear Models}
\index{surface ! deformations}
\index{multivariate general linear model}
\index{general linear model ! multivariate}

The multivariate general linear models (MGLMs) have been also used in modeling multivariate imaging features on brain surfaces. These models generalize {a widely used univariate GLM} by incorporating vector valued responses and explanatory variables \citep{anderson.1984, friston.1995.MGLM, worsley.1996, worsley.2004, taylor.2008, chung.2012.CNA}. {Hotelling's $T^2$ statistic} is a special case of MGLM that has been used primarily for inference on surface shapes, deformations and multi-scale features \citep{cao.1999.stat,chung.2012.CNA,gaser.1999,joshi.1998,kim.2012.NIPS,thompson.1997}. An example of this approach is {Hotelling's $T^2$-statistic} applied in determining the 3D brain morphology of HIV/AIDS patients \citep{lepore.2006}. {Hotelling's $T^2$-statistic}  is also applied to 2D deformation tensor at each mesh vertex on the hippocampal surface as a way to characterize AlzheimerÕs diseases \citep{wang.2011}.

Suppose there are a total of $n$ subjects and $p$ multivariate features of interest at each voxel. 
For MGLM to work, $n$ should be significantly larger than $p$. Let ${\bf J}_{n \times p} = (J_{ij})$ be the measurement matrix, where $J_{ij}$ is the measurement for subject $i$ and for  the $j$-th feature. The subscripts denote {the dimension of the matrix}. All the measurements over subjects for the $j$-th feature {are} denoted as ${\bf x}_j=(J_{1j}, \cdots, J_{nj})'$. The measurement vector for the $i$-th subject is denoted as ${\bf y}_i = (J_{i1}, \cdots, J_{ip})'$. ${\bf y}_i$ is expected to be distributed identically and independently over subjects. Note that 
$${\bf J} = ({\bf x}_1, \cdots, {\bf x}_p) = ({\bf y}_1, \cdots, {\bf y}_n)'.$$ 
We may assume the  covariance matrix of ${\bf y}_i$ to be
$$\mbox{Cov}({\bf y}_1) = \cdots = \mbox{Cov}({\bf y}_n) = {\bf \Sigma}_{p \times p} = (\sigma_{kl}).$$
With these notations, we now set up the following MGLM at each mesh vertex:
\bqn {\bf J}_{n \times p} = {\bf X}_{n \times k}{\bf B}_{k \times p} + {\bf Z}_{n \times q}{\bf G}_{q \times p} + {\bf U}_{n \times p}{\bf \Sigma}_{p \times p}^{1/2}.\label{eq:multivariate}\eqn
$\bf X$ is the matrix of contrasted explanatory variables while $\bf B$ is the matrix of unknown coefficients. Nuisance covariates are in matrix $\bf Z$ and the corresponding coefficients are in matrix $\bf G$. The components of Gaussian random matrix $\bf U$ are independently distributed with zero mean and unit variance. Symmetric matrix $\bf \Sigma^{1/2}$ is the square root of the covariance matrix accounting for the spatial dependency across different voxels. In MGLM (\ref{eq:multivariate}), we are usually interested in testing  hypotheses
$$H_0: {\bf B} = 0.  \; \mbox{\em vs.} \; H_1: {\bf B} \neq 0.$$ 
The parameter matrices in the model are estimated via the least squares method. The multivariate test statistics such as  Lawley-Hotelling trace or Roy's maximum root are used to test the significance of ${\bf B}$. When there is only one voxel, i.e. $p=1$, these multivariate test statistics collapse to  Hotelling's $T^2$ statistic \citep{worsley.2004, chung.2012.CNA}.

\subsection{Small-$n$ Large-$p$ Problems}
\index{small-$n$ large-$p$ problem}

GLM are usually fitted in each voxel separately. Instead of fitting GLM at each voxel, one may be tempted to fit the model in the whole brain surface.  For FreeSurfer meshes, we need to fit GLM over  300000 vertices, which causes the {\em small-$n$ large-$p$ problem} \citep{friston.1995.MGLM,schafer.2005,chung.2013.SCM}. 

Let ${\bf y}_j$ be the measurement vector at the $j$-th vertex. Assume there are $n$ subjects and total $p$ vertices in the surface. We have the same design matrix ${\bf Z}$ for all $p$ vertices. Then we need to estimate the parameter vector ${\boldsymbol \lambda}_j$ in
\bqn {\bf y}_j = {\bf Z} {\boldsymbol \lambda_j} \label{eq:normal-j} \eqn
for each $j$. Instead of solving (\ref{eq:normal-j}) separately at each vertex, we combine all of them together so that we have matrix equation
\bqn \underbrace{[{\bf y}_1, \cdots, {\bf y}_m]}_{\bf Y} = {\bf Z} \underbrace{[{\boldsymbol \lambda}_1, \cdots, {\boldsymbol \lambda}_m]}_{\boldsymbol \Lambda}\label{eq:normal-j2}.\eqn
The least squares estimation of the parameter matrix ${\boldsymbol \Lambda}$ is given by
$$\widehat{ \boldsymbol \Lambda} = ({\bf Z}'{\bf Z})^{-1}{\bf Z}' {\bf Y}.$$
Note that ${\bf Z}$ is of size $n$ by $p$ and ${\bf Z}'{\bf Z}$ is only invertible when  $n \ll p$. The least squares estimation does not provide robust parameter estimates for  $n \ll p$, which is the usual case in surface modeling. 
For small-$n$ large-$p$ problem, GLM need to be regularized using the L1-norm penalty \citep{banerjee.2006,banerjee.2008,chung.2013.SCM,friedman.2008,huang.2010,mazumder.2012}.

\subsection{Longitudinal Models}
\index{longitudinal}

So far we have only dealt with {an imaging data set} where the parameters of the model are fixed and {do} not vary across subjects and scans. Such fixed-effects models are inadequate in modeling the within-subject dependency in longitudinally collected imaging data. However, mixed-effects models can explicitly model such dependency \citep{fox.2002, milliken.2000, molenberghs.2005, pinheiro.2002}. There are three advantages of the mixed-effects model over the usual fixed-effects model. It explicitly models individual growth patterns {and} accommodates an unequal number of follow-up image scans per subject and unequal time intervals between scans. 

The longitudinal outcome $Y_i$  from the $i$-th subject is modeled using the mixed-effects model  \citep{milliken.2000} as
\bqn Y_i = X_i\beta + Z_i\gamma_i + e_i,\label{eq:surface-mixed}\eqn
where $\beta$ is the fixed-effects shared by all subjects. $\gamma_i$ is the {subject-specific} {random-effects} and $e_i \sim N(0,\sigma^2)$ is independent and identically distributed noise. $X_i$ and $Z_i$ are the design matrices corresponding to the fixed and random effects respectively for the $i$-th subject. 
We assume $\gamma_i \sim N(0, \Gamma)$ and $\epsilon_i \sim N(0, \Sigma_i)$ with some covariance matrices $\Gamma$ and $\Sigma_i$. Hierarchically we are modeling \eqref{eq:surface-mixed} as 
$$Y_i | \gamma_i \sim N(X_i \beta + Z_i \gamma_i , \Sigma_i), \; \gamma_i \sim N(0, \Gamma).$$ 
$\Gamma$ accounts for covariance among random effect terms. The within-subject variability between the scans is expected to be smaller than between-subject variability and explicitly modeled by $\Sigma_i$. The covariance of $\gamma_i$ and $\epsilon$ are expected to have block diagonal structure such that there is no correlation among the scans of different subjects while there is high correlation between the scans of the same subject:
$$ \mathbb{V}
\left(
\begin{array}{c}
  \gamma_i  \\
    \epsilon_i \\
\end{array}
\right) = 
\left(
\begin{array}{ccc}
 \Gamma & 0    \\
 0  & \Sigma_i     \\
\end{array}
\right).
$$
Subsequently, the {overall} covariance of $Y_i$ is given by 
$$\mathbb{V} Y_i = Z_i \Gamma Z_i' + \Sigma_i.$$
The random-effect contribution is $Z_i \Gamma Z_i'$ while the within-subject contribution is $\Sigma_i$.

\index{estimation ! restricted maximum likelihood}
The parameters and the covariance matrices can be estimated via the restricted maximum likelihood (REML) method \citep{fox.2002, pinheiro.2002}. The most widely used tools for fitting the mixed-effects model are the {\tt nlme} library in the {\tt R} statistical package \citep{pinheiro.2002}. However, there is no need to use {\tt R}  to fit the mixed-effects model. Keith Worsley has implemented the REML procedure in the {\tt SurfStat} package\footnote{The MATLAB package is available at {\tt www.math.mcgill.ca/keith/surfstat}.} \citep{chung.2012.CNA,worsley.2009}.

Here we briefly explain how to set up a longitudinal mixed-effect model in practice. 
In the usual fixed-effect model,  we have a linear model containing the fixed-effect term ${\tt age}_{i}$ for the $i$-th subject:
\bqn y_{i} = \beta_0 + \beta_{1} {\tt age}_{i}  
+ \epsilon_{i}, \label{eq:fixed1} \eqn
where $\epsilon_{i}$ is assumed to follow independent Gaussian. 
In (\ref{eq:fixed1}), every subject has identical growth trajectory $\beta_0 + \beta_1 {\tt age}$, which is unrealistic. Biologically, each subject is expected to have its own unique growth trajectory. So we assume each subject to have its own intercept $\beta_0 + \gamma_{i0}$ and slope $\beta_1 + \gamma_{i1}$:
\bqn y_{i}= \beta_0 + \gamma_{i0} + (\beta_1 + \gamma_{i1}) {\tt age}_{i}  
+ \epsilon_{i}. \label{eq:fixed2} \eqn
It is reasonable to assume the random vector $\gamma=(\gamma_{i0}, \gamma_{i1})$ to be multivariate normal. The model (\ref{eq:fixed2}) can be decomposed into fixed- and random-effect terms:
\bqn y_{i} = (\beta_0 + \beta_1 {\tt age}_{i} )+  (\gamma_{i0} + \gamma_{i1} {\tt age}_{i}) + \epsilon_{i}. \label{eq:random1} \eqn \\
The fixed-effect term $\beta_0 + \beta_1 {\tt age}_{i}$ models the linear growth of the population while the random-effect term $\gamma_{i0} + \gamma_{i1} {\tt age}_{i}$ models the subject specific growth variations. Incorporating additional factors and interaction terms are done similarly. 

\index{linear models ! random effects}
\index{longitudinal analysis }

\begin{figure}
\centering
\includegraphics[width=1\linewidth]{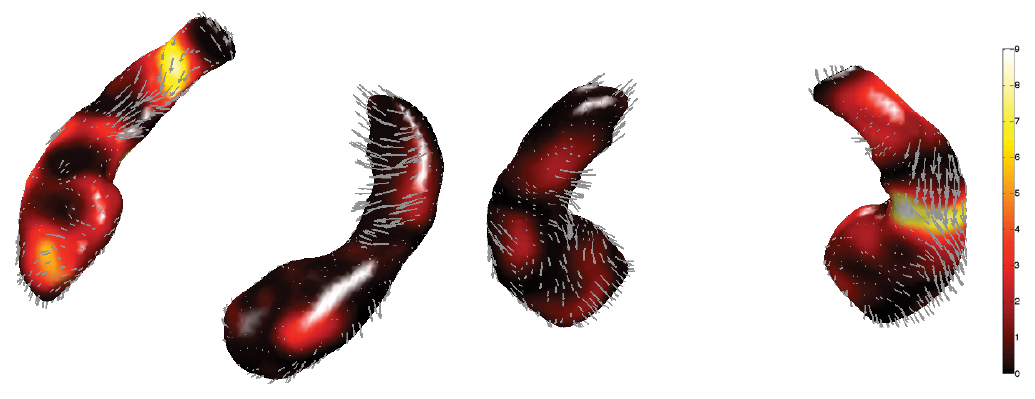} 
 \caption{$F$-statistics maps on testing the interaction between the income level and age while controlling for gender in a linear mixed-effects model. The arrows show the deformation differences  between the groups (high income - low income). Significant regions are only found in the tail and midbody regions of the right hippocampus.}
\label{fig:hippo-random}
\end{figure}

In the hippocampus case study, the first MRI scans are taken at $11.6 \pm 3.7$ years while the second scans are taken at $14 \pm 3.9$ years. We are interested in determining the effects of income level {on the shape of the hippocampus}. In Section \ref{sec:GLM}, we treated the second scans {as if they came from} independent subjects and modeled them using the fixed-effects model. Now we explicitly incorporate the dependency of repeated scans of the same subject. It is necessary to explicitly model the within-subject variability that is expected to be smaller than between-subject variability.  This can be done by introducing a random-effect term. The resulting F-statistic maps are given in Figure \ref{fig:hippo-fstat}-right. However, we did not detect any region that is affected {by income}. Thus, we tested the age and income interaction and found the regions of strong interaction (Figure \ref{fig:hippo-random}).

\begin{figure}[t]
\centering
\includegraphics[width=1\linewidth]{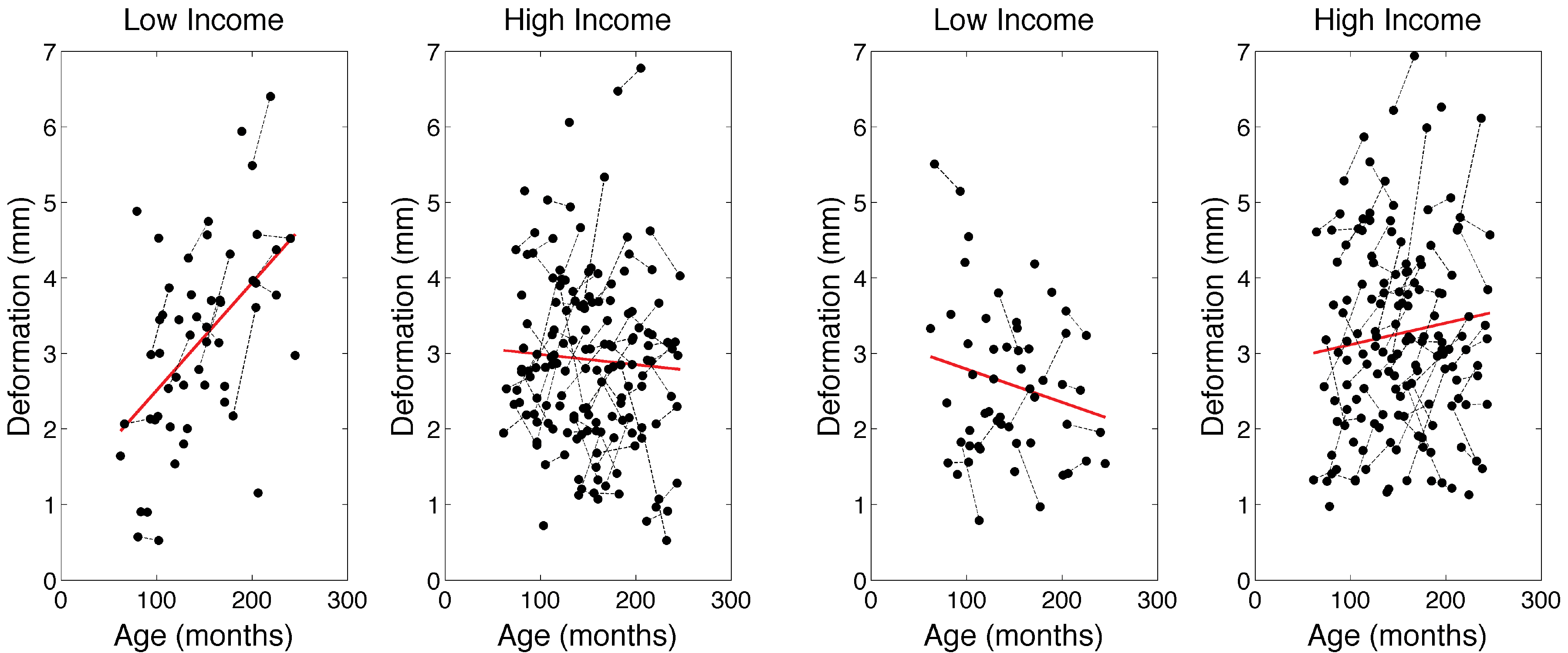} 
 \caption{The plots showing income level dependent growth differences in the posterior (left) and midbody (right) regions of the right hippocampus. The red lines are the linear regression lines. Scans within a subject {are} identified by dotted lines.}
\label{fig:hippo-plot}
\end{figure}

\subsection{Random Field Theory}
\index{random field}
\index{multiple comparisons}

Since we need to set up a GLM on every mesh vertex, it becomes a {\em multiple comparisons} problem. Correcting for multiple comparisons is crucial in determining overall statistical significance in correlated test statistics over the whole surface. For surface data, various methods are proposed: Bonferroni correction, random field theory \citep{ worsley.1994, worsley.1996}, false discovery rates (FDR) \citep{benjamini.1995, benjamini.2001, genovese.2002}, and permutation tests \citep{nichols.2002}. Among many techniques, the random field theory is probably the most natural in relation to surface data smoothing since it is able to explicitly control the amount of smoothing.

The generalization of a continuous stochastic process in $\mathbb{R}^n$ to a higher dimensional abstract space is called a {\em random field} \citep{adler.2007,chung.2012.CNA, dougherty.1999,joshi.1998,yaglom.1987}. In the random field theory \citep{worsley.1994, worsley.1996}, measurement $Y$ at position $x \in \mathcal{M}$ is modeled  as
$$Y(x) = \mu(x) + \epsilon(x),$$
where $\mu$ is the unknown signal to be estimated and $\epsilon$ is the
measurement error. The measurement error at each fixed $x$ can be modeled as a
random variable. Then the collection of random variables $\{\epsilon(x): x \in \mathcal{M} \}$ is called a {\em random field} or stochastic process. A measure-theoretic definition is given in \citep{adler.2007}. 

Detecting the regions of statistically
significant $\lambda(x)$ for some $x \in \mathcal{M}$ can be done via
thresholding the maximum of a random field defined on the
cortical surface \citep{worsley.1996, worsley.1999}. For instance, $T$ random field on the surface $\mathcal{M}$ is
defined as
$$T(x) = \sqrt{n}\frac{M(x)}{S(x)},$$
where $M$ and $S$ are the sample mean and standard deviation over the
$n$ subjects. Under the null hypothesis
$$H_0: \mu( x)=0 \text{ for all } x \in \mathcal{M},$$
$T(x)$ is distributed as a student's $T$ with $n-1$ degrees of freedom at each voxel $
x$. The pvalue of the local maxima of the $T$ field will give
a conservative threshold compared to FDR \citep{worsley.1996}.

For sufficiently high threshold $y$, we can show that \bqn P\Big(
\max_{x \in
\mathcal{M}} T({\bf x}) \geq y\Big) \approx
\sum_{i=0}^3 \phi_i(\mathcal{M})\rho_i(y),\eqn where
$\rho_i$ is the $i$-dimensional EC-density and the Minkowski
functional $\phi_i$ are
$$\phi_0(\mathcal{M})=2, \;
\phi_1(\mathcal{M})=0, \;
\phi_2(\mathcal{M}= \|\mathcal{M} \|, \;
\phi_3(\mathcal{M})=0$$ and $\| 
\mathcal{M}\|$ is the total surface area of $
\mathcal{M}$ (Worsley, 1996a). When diffusion or heat kernel smoothing with
given FWHM is applied on surface $\mathcal{M}$, the 0-dimensional and 2-dimensional EC-density
becomes
$$\rho_0(y)=\int_y^{\infty}
\frac{\Gamma(\frac{n}{2})}{((n-1)\pi)^{1/2}\Gamma(\frac{n-1}{2})}
\Big(1+\frac{y^2}{n-1}\Big)^{-n/2}\; dy,$$
$$\rho_2(y)=\frac{1}{\text{FWHM}^2}\frac{4 \ln 2}{(2\pi)^{3/2}}\frac{\Gamma(\frac{n}{2})}{(\frac{n-1}{2})^{1/2}
\Gamma(\frac{n-1}{2})}y\Big(1+\frac{y^2}{n-1}\Big)^{-(n-2)/2}.$$

The excursion probability, which is the probability of obtaining false positives for the one-sided alternate hypothesis,  is approximated by the following formula:
$$P\Big( \max_{x \in \mathcal{M}} T(x)
\geq y\Big) \approx 2\rho_0(y) + \|\mathcal{M}
\|\rho_2(y).$$  
For smoothing cortical thickness \citep{chung.2012.CNA,chung.2005.ipmi}, an FWHM of between 20 to 30 mm is recommended.  This FWHM reflects the spatial frequency associated with the sulcal pattern. For measurements on smaller subcortical structures such as hippocampus and amygdala, significantly smaller FWHM is recommended.  For amygdala and hippocampus, 0.5--1 mm would be sufficient. 

In the hippocampus case study, we did not detect any statistically significant group difference at 0.01 level after correcting for multiple comparisons in both the left and right hippocampi. However, we obtained highly focalized regions of group difference in the growth rate, the interaction between income level and age, in the right hippocampus (corrected pvalue = 0.03). The posterior region is enlarging while the midbody and the anterior parts are shrinking in children from low-income families (Figure \ref{fig:hippo-random} and \ref{fig:hippo-plot}). On the other hand, the pattern is opposite for children from high-income families. 
Note that the right hippocampus is involved in the active maintenance of associations with spatial information \citep{piekema.2006}. Future studies investigating the relationship between family socioeconomic status and spatial information processing measures are warranted.

\section*{Acknowledgements}
This study was supported by NIH grants UL1TR000427, MH61285 and MH68858, EB022856 and EB028753, and NSF grant MDS-2010778.
\bibliographystyle{agsm} 
\bibliography{reference.2022.03.13}

\end{document}